\documentclass[sigconf, nonacm]{acmart}

\newcommand\vldbyear{2024}
\newcommand\vldbworkshop{7th International Workshop on Big Data Visual Exploration and Analytics
 (BigVis 2024)}
\newcommand\vldbauthors{\authors}
\newcommand\vldbtitle{\shorttitle}
\newcommand\vldbavailabilityurl{}
\newcommand\vldbpagestyle{empty}

\usepackage{listings}
\usepackage{dsfont}

\usepackage{amsgen}
\usepackage{subcaption}
\usepackage{cleveref}

\usepackage{bm}
\usepackage{xcolor,colortbl}
\usepackage{pifont}
\usepackage{times}%
\usepackage{wrapfig}
 \usepackage[justification=centering]{caption}
\usepackage{color}
\usepackage{multirow}
 \usepackage{graphicx}
\usepackage[linesnumbered, vlined, ruled,algo2e]{algorithm2e}
\usepackage{booktabs}
\usepackage{amsfonts}
\usepackage{amsmath}
\usepackage{mathtools}
\usepackage{amsthm}
\usepackage{soul}
\usepackage{epstopdf}
\usepackage{url}
\usepackage{setspace}
 \usepackage{hyperref}
\hypersetup{
    colorlinks,%
    citecolor=black,%
    filecolor=black,%
    linkcolor=black,%
    urlcolor=black,%
}

\graphicspath{{figures/}{plots/}}
\definecolor{dark-gray}{gray}{0.2}
\newcommand{\myFontP}[1]{{\fontfamily{ppl}\selectfont #1}} %Adobe Palantino
\newcommand{\RED}[1]{  {\color{red}{#1}}}

\newif\iftodo

\newif\ifcheck

\newcommand{\eat}[1]{}

 \newcommand{\T}{\ensuremath\mathcal{T}}

 \newcommand{\fnc}[1]{\scalebox{0.9}{\textsf{#1}}}

\DeclareMathOperator*{\argmin}{arg\,min}

\renewcommand{\qed} {\hfill$\blacksquare$}

\eat{

}
\newcommand{\stitle}[1]{\bigskip\noindent\textbf{#1}} %5.5 pt
\newcounter{prob}

\newcounter{theorcnt}

\newcounter{ex}

\usepackage{tabularx}

\SetKw{Logand}{and}
\SetKw{Logor}{or}
\SetKw{Lognot}{not}

\SetKw{mybreak}{break}	%define your own keywords
\SetKw{mycontinue}{continue}	%define your own keywords

\SetKwInput{KwVar}{Variables}
\SetKwInput{KwPar}{Parameters}

\SetKwData{fnp}{\textsf{findFileTiles}}
\SetKwData{csp}{\textsf{splitRequired}}

\SetKwData{myempty}{empty}
\SetKwBlock{myblock}{beginnikos}{end} %define your own keywoed blocks
\SetAlgoInsideSkip{}
\SetAlgoSkip{medskip}
\SetVlineSkip {1.5mm}
\SetAlgoCaptionSeparator{.}

\newcommand{\mycomment} [1] 	{\tiny{\textcolor{dark-gray}{\myFontP{{#1}}}}} %\textcolor{dark-gray}
\SetKwComment{Comment}{\mycomment{/\!\!/}}{}
 \DontPrintSemicolon
\SetAlgoCaptionLayout{small}
\SetProcNameSty{small}
\SetProcArgSty{small}

\hypersetup{
	colorlinks,%
	citecolor=black,%
	filecolor=black,%
	linkcolor=black,%
	urlcolor=black,%
	pdftitle={Partial Adaptive Indexing for Approximate Query Answering},
	pdfauthor={Stavros Maroulis, Nikos Bikakis,  Vassilis Stamatopoulos, George Papastefanatos},
	pdfsubject={Adaptive Indexing for Approximate Query processing},
	pdfkeywords={Big Data visualization tools systems, query Approximation, adaptive indexing, Big Data Visual Analytics, In-situ Query Processing, RawViz, Learning indexes, Steerable Progressive Visual Exploration, Big Raw Data, User driven Incremental Processing, In-Situ Visual Analytics, Interactive indexes, Progressive and Adaptive indexing, Information visualization, HCI, Data Exploration}
}

\renewcommand{\baselinestretch}{0.94}

\begin{document}

\title{Partial Adaptive Indexing for Approximate Query Answering}

% \title{Approximate Interactive In-Situ Visual Analysis through Partial Adaptive Indexing[Work-in-progress]}

% Towards Adaptive Indexing for Approximate Visual Analytics [Work-in-progress]}
 %Query Answering}

\author{Stavros Maroulis}
\affiliation{%
  \institution{ATHENA Research Center, Greece}
  }

 \author{Nikos Bikakis}
\affiliation{%
  \institution{Hellenic Mediterranean University \& \\ ATHENA Research Center, Greece}
  }

\author{Vassilis Stamatopoulos}
\affiliation{%
  \institution{ATHENA Research Center, Greece}
  }

\author{George Papastefanatos}
\affiliation{%
  \institution{ATHENA Research Center, Greece}
  }

\begin{abstract}
In data exploration, users need to  analyze large data files quickly, aiming to minimize   data-to-analysis time. While recent adaptive indexing approaches address this need, they are cases where demonstrate poor performance. Particularly, during the initial queries, in regions with a high density of objects, and in very large files over commodity hardware.
This work introduces an approach for adaptive indexing driven by both query workload and user-defined accuracy constraints to support approximate query answering. The approach is based on partial index adaptation which reduces the costs associated with reading data files and refining indexes. We leverage a hierarchical tile-based indexing scheme and its stored metadata to provide efficient query evaluation, ensuring accuracy within user-specified bounds. Our preliminary evaluation demonstrates improvement on  query evaluation time, especially during initial user exploration.

% Future work will extend this approach to support categorical-based group-by aggregations and integrate it into progressive visualization environments for improved performance and memory management.

\end{abstract}

\maketitle

%%% do not modify the following VLDB block %%
%%% VLDB block start %%%
\pagestyle{\vldbpagestyle}
\begingroup\small\noindent\raggedright\textbf{VLDB Workshop Reference Format:}\\
\vldbauthors. \vldbtitle. VLDB \vldbyear\ Workshop: \vldbworkshop.\\ %\vldbvolume(\vldbissue): \vldbpages, \vldbyear.\\
%\href{https://doi.org/\vldbdoi}{doi:\vldbdoi}
\endgroup
\begingroup
\renewcommand\thefootnote{}\footnote{\noindent
This work is licensed under the Creative Commons BY-NC-ND 4.0 International License. Visit \url{https://creativecommons.org/licenses/by-nc-nd/4.0/} to view a copy of this license. For any use beyond those covered by this license, obtain permission by emailing \href{mailto:info@vldb.org}{info@vldb.org}. Copyright is held by the owner/author(s). Publication rights licensed to the VLDB Endowment. \\
\raggedright Proceedings of the VLDB Endowment. %, Vol. \vldbvolume, No. \vldbissue\ %
ISSN 2150-8097. \\
%\href{https://doi.org/\vldbdoi}{doi:\vldbdoi} \\
}\addtocounter{footnote}{-1}\endgroup
%%% VLDB block end %%%

%%% do not modify the following VLDB block %%
%%% VLDB block start %%%
\ifdefempty{\vldbavailabilityurl}{}{
\vspace{.3cm}
\begingroup\small\noindent\raggedright\textbf{VLDB Workshop Artifact Availability:}\\
The source code, data, and/or other artifacts have been made available at \url{\vldbavailabilityurl}.
\endgroup
}
%%% VLDB block end %%%

 \vspace{-6pt}

\section{Introduction}
\label{sec:intro}

In data exploration, users often need to \textit{interact with and analyze} large data files without the hassle of full-fledged database configuration and lengthy data loading and indexing times. A common objective in such scenarios is to minimize the \textit{data-to-analysis time} while ensuring efficient visual exploration and analytical operations.

In such exploration scenarios, users might not always \textit{require exact results}. There are interactive scenarios where response time is more crucial than result accuracy \cite{hal-04361344, KimBPIMR15, 0001S20, ParkCM16, RahmanAKBKPR17, liu2013immens}. Several visual analytic tasks, such as class or outlier analysis in scatterplots, pair-wise comparison of spatial areas on maps, usually start with approximate  aggregated insights, which can be used by the experts to quickly identify specific areas in the exploration space for further analysis. Approximate query processing (AQP) is a long-studied problem in the areas of databases and information visualization; however, there is a missing gap when AQP is coupled with the \textit{in-situ paradigm}.

\textit{In-situ paradigm} has been adopted for data exploration, enabling on-the-fly analysis of large raw data sets such as CSV or JSON files \cite{MaroulisS22, IS, Alagiannis2012, Olma20}. In-situ techniques aim to bypass the overhead of fully loading and indexing data in a DBMS while offering efficient query evaluation over the raw data files. Previous works on the visual exploration of raw data files \cite{MaroulisS22, IS} have focused on building adaptive indexes over the raw data objects. These methods leverage the locality-based behaviour of visual exploration, in which the user explores a specific area in the data space, gradually widening the analysis in neighboring areas.  Thus, they seek to minimize index initialization time by initially creating a "crude" version of the index, e.g., around the initial area of interest, dynamically extending and adapting it based on user exploration.

However, since the initial index is only a "crude" version, \textit{initial queries can be slower as the index adapts and refines based on user exploration}. Additionally, \textit{for very large raw data files or regions with a high density of objects}, even an index that has sufficiently adapted can result in significant query times, thus compromising interactivity.

 Considering these challenges, \textit{our goal is to reduce response time by providing approximate   results and performing ``partial'' index adaptation}. Partial index adaptation
 allows us to reduce the costs associated with reading the raw data file (i.e., I/O cost) and refining the index, e.g., recomputing metadata, reorganizing objects.

\vspace{-6pt}
\stitle{Contribution.}
In this paper we provide our ongoing work and preliminary results on the problem of \textit{adaptive indexing driven by both the query workload and the accuracy constraints set by the user}.
Particularly, \textit{given an accuracy constraint we attempt to reduce the response time, by performing
a minimum number of adaptations (and consequence the I/O's) such as the result's accuracy is larger than a threshold}.
The proposed methods are going to be integrated to our
 \textit{RawVis framework}\footnote{The source code is available at \textit{\href{https://github.com/VisualFacts/RawVis}{https://github.com/VisualFacts/RawVis}}} \cite{MaroulisS22,MaroulisBPVV21,IS}, supporting approximate query answering via partial index adaptation.

  \vspace{-6pt}
\stitle{Related Work.}
In visual analytics, \textit{approximate processing} techniques (a.k.a.\ \textit{data reduction}), such as sampling and binning, have been used to improve efficiency and address the visual information overloading problem \cite{AQP++,0001S20, KimBPIMR15, ParkCM16, liu2013immens}.
Recent works also consider visualization parameters to ensure perceptually similar visualizations and offer visualization-aware error guarantees \cite{MaroulisVLDB24}.
Further, \textit{progressive visualization} approaches perform computations in small, incremental steps, providing users with progressively improving approximate results \cite{OM323, hal-04361344,Davos21, RahmanAKBKPR17, AgarwalMPMMS13}. %FeketeF0S18,FisherPDs12, AngeliniSSS18
In contrast, we focus on implementing an adaptive indexing scheme for approximate query answering.
\textit{Adaptive indexing} techniques dynamically adjust indexes based on query workloads to optimize data access  \cite{MaroulisS22,IS, ZardbaniMIK23,0002ZMK23,HolandaM21,Olma20, GhoshEJ19, Alagiannis2012}. %ZardbaniAK20
Compared to our work, these studies are based on exact query answering.

Recent techniques for AQP, which combine pre-computed aggregates and data sampling to facilitate interactive analysis, are similar to our work but in a different context \cite{AQP++, Liang21}. However, these approaches neither utilize adaptive indexing to support efficient query evaluation over large data files nor dynamically adjust the granularity of their pre-computed aggregates to accommodate the dynamic nature of data exploration scenarios. In contrast, our work leverages an adaptive index and pre-computed aggregates to perform in-situ adaptive indexing with error-bound guarantees. Additionally, we allow for ``partial'' index adaptation to reduce associated costs while considering the user's accuracy constraints.

\section{Framework Overview}
\label{sec:method}

\begin{figure*}[t]
\vspace{-35pt}
	\centering
\hspace*{-10pt}	\includegraphics[width=5.9in]{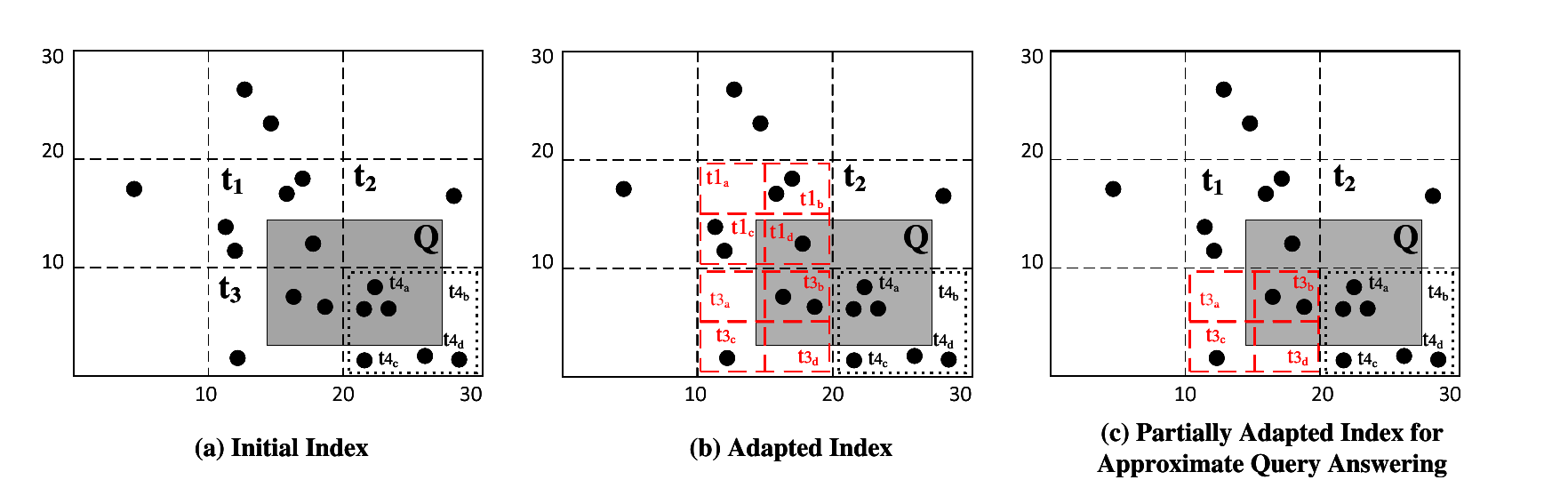 }
 \vspace{-12pt}
\caption{Index Adaptation Example (a) Initial index structure; (b) Exact query answering, splitting tiles $t_1$ and $t_3$; (c) Approximate query answering, splitting only $t_3$ and providing results within user accuracy constraints}
 \label{fig:example}
 \vspace{-4pt}
 \end{figure*}

\subsection{Exploration Model}
\label{sec:scenario}

In our scenario, we consider a user visually exploring data stored in a file (e.g., CSV file) using a 2D visualization technique (e.g., map, scatter plot), and analyzing it using visual tools (e.g., bar and line charts, heatmaps) as well as aggregate and statistical measures (e.g., mean, Pearson correlation) \cite{MaroulisBPVV21}. The data attributes may be numeric, spatiotemporal, categorical, or textual. At least two of these attributes must be numeric (e.g., longitude, latitude) and can be mapped to the X and Y axes of the 2D visualization (\textit{axis attributes}).

Our framework is based on a formal \textit{exploration model} that defines a set of exploratory and analytic operations to formulate user interactions \cite{MaroulisS22}. It implements basic \textit{exploration operations} over a 2D visualization plane, such as pan, zoom, filter, view object details, and selection of objects by defining 2D areas (i.e., range queries) over the visualized objects. Beyond exploratory operations, the model defines \textit{analytic operations} that allow the user to visually analyze the objects by generating several types of visualizations that can aggregate, compare, or provide statistics on data properties.

\subsection{Indexing Scheme}
\label{sec:indx}

For the sake of simplicity, we describe the basic idea of our methods over the hierarchical
tile-based VALINOR index (referred as \textit{index}) \cite{IS}, which is a simplified
version of our VETI index \cite{MaroulisS22}.%
\footnote{Compared to VALINOR, the VETI index combines tiles with tree structures enabling categorical-based aggregations, and supports \mbox{resource-aware} index management.}

The index is stored in main memory and organizes the data objects into
hierarchies of non-overlapping rectangle tiles. The index’s tiles
are defined over the domains of axis attributes.
Each tile is associated with \textit{metadata} (e.g., average, sum, count) that enables statistics computations.
Particularly, metadata are  numeric values calculated
by algebraic aggregate functions over one or more attributes of
the objects.

Consider the example where the 2D canvas is a map, the data points are  hotels with ratings and their location on the map is given by the longitude and latitude values. Figure~\ref{fig:example}~(a) presents an example of the basic structure of an index, which divides the 2D
space into $3 \times 3$ equally sized disjoint tiles
($t_1$$\sim$$t_4$),
where the tile $t_4$ is further divided into $2 \times 2$ subtitles ($t_{4_a}$$\sim$$t_{4_d}$), forming a tile hierarchy\footnote{For simplicity here we present equally sized tiles.}. Each tile has a range on the longitude and latitude attributes, contains the data points within that range and keeps an aggregate value (e.g., average rating) for the containing set.

\stitle{Index Initialization.}
 Initially, a “crude”, lightweight initial version of the index is built, and progressively
adjusts itself to the user interactions, by splitting the tiles visited into more fine-grained ones.

\stitle{Index Adaptation.}
In what follows, we outline the index adaptation used for exact query answering.
RawVis employs a progressive index adaptation technique that attempts to reduce the I/O
operations and computations by adjusting the index based on the user interactions, e.g.,
exploration areas and required statistics.
Adaptation is performed by modifying the structure of the index
(e.g., tile size); and by enriching and updating
its “information” (e.g., metadata).
The adaptation method incrementally splits the tiles that overlap
with a query into smaller subtiles. The splitting process considers
factors related to I/O cost in order to decide whether to perform a
split or not.
Considering the locality-based characteristics of the
exploration scenarios, tile splitting increases the likelihood that a
future query will fully overlap a tile in the area, which the user exploration
focuses on.
The case of fully overlapped tiles allows the index
to use the existing metadata, improving the query performance by
reducing I/O operations on the file.
In conjunction with the tile splitting, the index may be enriched by computing different metadata.

Figure~\ref{fig:example}~(b) shows the index adaptation when query $Q$ is posed.
The tiles $t_1$ and $t_3$ that overlap with $Q$ are split to $2 \times 2$ subtiles.
Note, for simplicity, we assume that there is no need to further split $t_{4_a}$$\sim$$t_{4_d}$ subtiles.
So, in this case we have to reorganize the objects included in $t_1$ and $t_3$ tiles, and compute the metadata for each subtile.

\section{Partial Index Adaptation for Approximate Query Answering}
\label{sec:adapt}

% \subsection{Introduction}
% Considering that there are exploration scenario where users might not  require exact results

Our goal is to support approximate query evaluation over the index, reducing the cost of reading raw data from the file and the cost of adapting the index while ensuring that the results' accuracy meets a given constraint.
While leveraging the index structure and the aggregate metadata can significantly reduce file accesses (i.e., execution time), there are cases such as the initialization phase and dense areas, where query execution exhibits lower performance.

The performance in the aforementioned cases become even more challenging  when the user's exploratory queries involve attributes that are not directly stored in the index.
Recall that,  the index stores the axis attributes (e.g., longitude, latitude) used in 2D visual explorations. This enables efficiently determining which objects fall within the query window without direct access to the raw data file.
However, queries involving analytic functions that utilize attributes not directly indexed (e.g., aggregating non-axis attributes) may necessitate file access to compute exact results.
% Therefore, we propose methods for approximate query evaluation and partial index adaptation to mitigate these challenges.

%
The  cases of exact and approximate query processing  are illustrated in Figure~\ref{fig:example}.
 We assume that the index is already initialized. Figure~\ref{fig:example}(a) depicts the index before the query $Q$ is posed, whereas Figure~\ref{fig:example}(b) and (c) depict the updated index after $Q$ evaluation with 100\% accuracy and the approximate case, respectively. Assume that the query $Q$ requests an aggregate function (e.g., average rating of the hotels) to be computed over all objects within $Q$ range.

\vspace{-6pt}
\stitle{Index Adaptation for Exact Query Answering Example.}
First, we identify the tiles that overlap with the query, i.e., $t_1$, $t_2$, $t_3$, and $t_{4_a} \sim t_{4_d}$. Next, we identify which of the overlapped tiles require file access. Tiles $t_2$ and $t_{4_b} \sim t_{4_d}$ are skipped, as they do not include any of the selected objects. Tile $t_{4_a}$ contains objects, but it is fully contained in the query and, its indexed metadata can be used to compute the aggregates requested for $Q$, without additional data access.

However, tiles $t_1$ and $t_3$ are partially contained in the query, i.e., their metadata concerns the entire tile and not an aggregate value for the specific selected objects requested for $Q$. Thus, we need to read from the file the required attributes' values of the objects in $t_1$ and $t_3$ that are within the query. This results in reading three objects. Simultaneously, we split the tiles into four sub-tiles and compute and store metadata for the newly created sub-tiles $t_{1_d}$ and $t_{3_b}$.
%further adapting the index structure.

\subsection{Approach Overview}

In this section we provide some basic concepts and the problem definition.

\vspace{-6pt}

% \vspace{-6pt}

\stitle{Query Confidence Interval.}
A key aspect is that for the partially contained tiles, we can compute (without accessing the file) the number of objects within the window query via the objects axis values stored in the index.
Using the number of objects (\textit{count}) along with the \textit{sum}, \textit{min}, and \textit{max} aggregate metadata stored in each tile, we can deterministically bound the aggregate value of the objects in the query.
This allows us to establish a \textit{query confidence interval} and guarantee that the actual aggregate value falls within it. Based on that, we can approximate various aggregate functions, such as \textit{sum}, \textit{mean}, \textit{min} and \textit{max}.

For example, assume that we wish  to compute the query confidence interval for the \textit{sum} function over a non-axis attribute \({A}\) for the objects within the window query $Q$.
In this case, the \textit{query confidence interval} for the \textit{sum} function is calculated by using the precomputed metadata  of each tile (i.e., objects count, minimum value, maximum value). Particularly, the \textit{query confidence interval} is defined as:

 \vspace{-12pt}

{\scriptsize{
\begin{align*}
   & \left[\sum_{t \in \T_{f}} \text{sum}_{{A}}(t) + \sum_{t \in \T_{p}} \text{count}(t \cap Q) \cdot \text{min}_{{A}}(t),\right. \\
&\left. \sum_{t \in \T_{f}} \text{sum}_{{A}}(t) + \sum_{t \in \T_{p}} \text{count}(t \cap Q) \cdot \text{max}_{{A}}(t) \right],
\end{align*}
}}

\noindent
where $\T_{f}$ denote the fully-contained tile set and $\T_{p}$ the partially-contained tiles set; $\text{sum}_{{A}}(t)$,  $\text{min}_{{A}}(t)$ and $\text{max}_{{A}}(t)$ the sum, the minimum and the maximum value of the attribute $A$ in tile $t$, respectively; and $\text{count}(t \cap Q)$ the number of objects inside the tile $t$ that are selected by the query $Q$.

This computation can be generalized to the \textit{mean}, \textit{min} and \textit{max}. For the \textit{mean}, the confidence interval is determined by dividing the sum query confidence interval by the total count of objects included in the query. For the \textit{min} and \textit{max} aggregates, the query confidence interval is derived by considering the fully contained tiles' min or max values respectively and bounding the partially contained tiles' values within their min-max range.

 \stitle{Tile Confidence Interval.}
Following a similar approach as the query confidence interval, confidence intervals can be computed for partially-contained tiles. For example, for the \textit{sum} aggregate function, given a query $Q$, a partially-contained tile $t$ and a non-axis attribute $A$, the \textit{tile confidence interval} for the \textit{sum} over $A$ is defined as:
$\left[\text{count}(t \cap Q) \cdot \text{min}_{{A}}(t), \text{count}(t \cap Q) \cdot \text{max}_{{A}}(t) \right]$.

\eat{
\vspace{-6pt}
{\scriptsize{
\begin{align*}
\left[\text{count}(t \cap Q) \cdot \text{min}_{{A}}(t), \text{count}(t \cap Q) \cdot \text{max}_{{A}}(t) \right]
\end{align*}
}}
 }

\stitle{Approximate Value}
For an aggregate function included in the query, we compute an \textit{approximate value} using: (1) the exact values from the query's fully contained tiles; and (2) approximate values from the query's partially contained tiles. For the partially contained tiles, the approximate values are derived using the tile's aggregate metadata. For example, the approximate value for the sum function is computed by multiplying each partially contained tile's mean value (derived from its min and max values) with the count of objects within the query range in that tile.

\vspace{-6pt}

\stitle{Upper Error Bound.}
By considering query confidence interval, we derive a relative \textit{upper error bound}.
The upper error bound is computed by normalizing the maximum difference between the approximate value computed and the query confidence interval bounds.
 \eat{
If this is not within the user-specified accuracy constraints, we need to read from the file the objects from the partially contained tiles  to improve the accuracy of the query result.
}

\stitle{Problem Definition.}
 Recall that, splitting is performed over the tiles that are partially contained within the query. When a split is performed we have to read from file the objects included in the partially contained tile and compute tile's metadata.

Let $\T$ be the \textit{set of query's partially contained tiles}.
Let  \fnc{process}$(t)$  be a function, which  \textit{processes a partially contained tile} $t \in \T$, i.e., splits the tile, reorganizes the objects in the subtiles, reads from the file the values of the objects included in $t$, and computes the metadata of each subtile.
Let \fnc{process}$(t).cost$  be the \textit{time required to process the tile} $t$.

Finally,  \fnc{error\_bound}$(\T, \T')$ denote \textit{the upper error bound} between the query answers resulting from processing all the partially contained tiles $\T$ (exact answer) and a subset $\T' \subseteq \T$ (approximate answer).

% \stitle{Problem.}
Given the set of query's partially contained titles $\T$, our problem is to find a set $\T' \subseteq \T$ such that the cost of processing the tiles $\T'$ is  minimized, and the answer error bound is lower or equal to a user-specified accuracy constraint $\phi$. Formally:

{\small{
	\begin{center}
	\hspace*{-5pt} $\underset{\substack{\T' \\  \T' \subseteq \T }}{\argmin} \:
{\underset{\forall t \in \T'}{ \sum}\fnc{process}(t).cost}$
		   s.t.\   \fnc{error\_bound}$(\T, \T') \leq \phi$
\end{center}
}}

% \vspace{-6pt}

\stitle{Processing Partially Contained Tiles.}
%
% The objects from the partially contained tiles are read at a tile-basis.
In order to find the subset of partially contained tiles which will process, we use the following simple approximation method.

 For each partially contained tile $t \in \T$ we compute \textit{a score} $s(t)$ that combines  factors related to accuracy and processing cost.
Particularly, for each $t$ we consider: (1) \textit{the width of the tile confidence interval} $w(t)$, that formulates the "degree of inaccuracy" of $t$. Note that, tiles with wider confidence intervals are considered more inaccurate; and
 (2) $\text{count}(t \cap Q)$ that is \textit{the number of objects inside $t$ that are selected by the query} $Q$, that formulates the processing cost.
 Formally, the score $s(t)$ of a tile $t$ is defined as:
 \mbox{$s(t)=\alpha \cdot w(t) + (1-\alpha) /\text{count}(t \cap Q)$},
  where $\alpha \in [0,1]$ formulates the trade-off between the two metrics.
Note that, $w(t)$ and $\text{count}(t \cap Q)$ are normalized to $[0,1]$.

We define a tiles selection policy that prioritizes the process of  the tiles from $\T$ with the largest scores, progressively refining our results until they meet the user-specified accuracy constraint. So, the processed tiles correspond to tiles set $\T'$.

For example, in Figure~\ref{fig:example}(c) we depict a partially adapted index  after evaluating the query $Q$. We have two partially contained tiles that include objects selected by the query, i.e., $t_1$ and $t_3$
Assume that the tile $t_3$ has larger score than $t_1$
% has a wider confidence interval than the tile $t_1$ and thus is less accurate.
So, we first process $t_3$, i.e., perform a splitting and we read from file the objects within tile.
Then, if the estimated value for the query falls within the user-defined error bounds, there is no need to access the file for the objects in the tile $t_1$.
So, the process of $t_1$ is avoided.

 \vspace{-6pt}

\section{Preliminary Evaluation}
\label{sec:eval}

In this section, we present a preliminary evaluation focusing on response time improvements achieved through partial index adaptation under different accuracy constraints. We used a sequence of queries and measured the evaluation time under $1$\% and $5$\% accuracy constraints compared to the exact query answering method. We used the synthetic dataset from \cite{IS, MaroulisS22} with 10 numeric columns (11 GB). Each query, defined over two numeric attributes, specifies a window containing approximately 100K objects and is shifted $10$$\sim$$20$\% randomly to simulate a map-based exploration path.
Finally, the score $s$ for each tile used by the tile selection policy considers only the width of the tile confidence interval, i.e., $\alpha = 1$.

% The dataset used corresponds to a synthetic CSV file of 100M \textit{data objects} having 10 columns (11 GB) with \textit{numeric attributes} following a uniform distribution. Each query is defined over two numeric attributes specifying a window size containing approximately 100K objects, similar to \cite{IS, MaroulisS22}.

\begin{figure}[t]
	\centering
\hspace*{-6pt}	\includegraphics[width=3.1in]{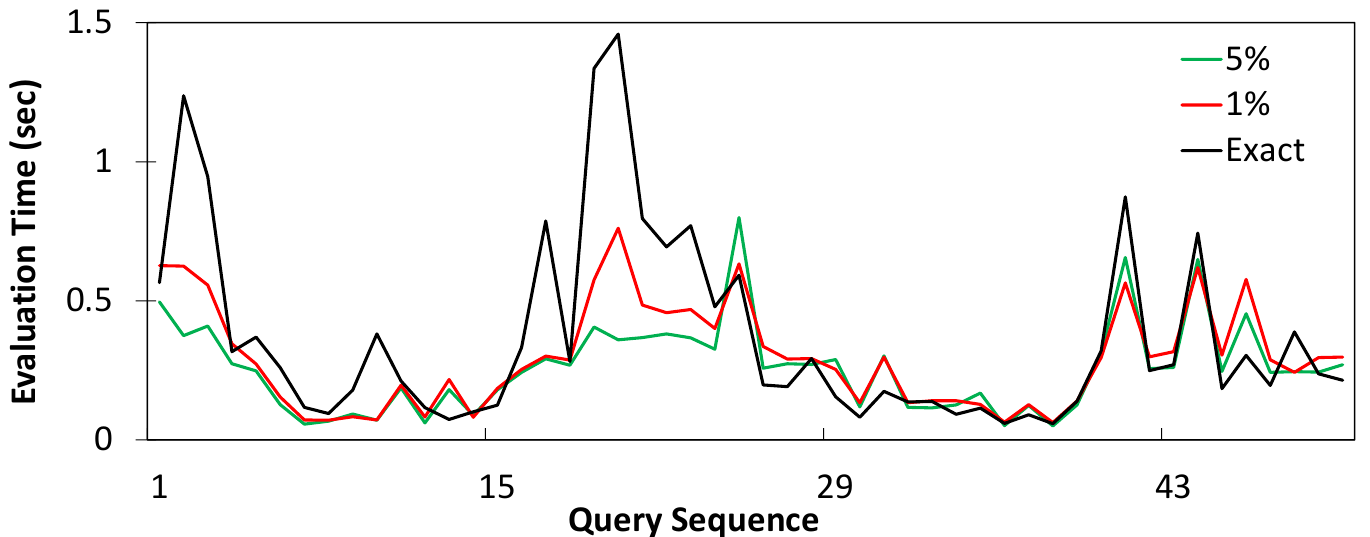 }
\vspace{-4pt}
    \caption{Evaluation Time   for Different Error Bounds}

\vspace{-4pt}
 \label{fig:plot}
 \end{figure}

Figure~\ref{fig:plot} illustrates the evaluation times for a sequence of 50 \textit{overlapping queries}.
The black line represents the evaluation time for exact query answering, while the red and green lines represent the evaluation times for $1$\% and $5$\% max error bounds, respectively.  \eat{Recall that the index stores the values for the axis attributes (e.g., latitude and longitude) for the objects indexed in it, while for the non-axis attributes, we need to evaluate their aggregate values for the window query using the aggregate metadata stored in each tile and reading the objects of the partially contained tiles from the data file.}
The peaks in the evaluation times can be attributed to the exploration of previously unvisited areas, where the index is less refined.

The results indicate that our partial index adaptation approach can   reduce query evaluation times, especially in the early stages of user interaction. The trade-off between accuracy and performance is evident, with higher accuracy constraints leading to slightly longer evaluation times but still outperforming the exact method.

The evaluation times closely follow the number of objects (i.e., CSV file rows) that need to be read from the raw data file. In the approximate cases, we can skip reading objects for some tiles partially overlapping the window query, while utilizing their aggregate metadata to bound their confidence interval. This is particularly effective for the first 20 queries where only a "crude" version of the index is available. As the index adapts for the exact evaluation scenario, the queries become comparable or slightly faster than the approximate ones because the index has been substantially adapted in the areas explored by the user, reducing the need to access the raw data file repeatedly. In contrast, the approximate cases may result in less efficient query evaluation as the index is not as thoroughly adapted.

Overall, the results seem promising, with most queries being evaluated faster for the approximate cases, especially in the first queries of the exploration scenario. For example, at query 20, the $5$\% accuracy method performs considerable well, being $4\times$ faster than the exact method, while the $1$\% accuracy method is about twice as fast.
However, in some queries, the approximate methods can result in higher evaluation times because the exact method has progressively refined the index more, allowing for quicker evaluations. This uncommon behavior highlights the trade-off between partial and full index adaptation, which we aim to improve by developing advanced tile selection policies and enabling more index adaptation even if the accuracy constraints have been satisfied.
Overall, considering the whole exploration scenario, the $5$\% and 1\% methods are about $40$\% and $30$\% faster, respectively.

\stitle{ACKNOWLEDGMENTS.}
This work was supported by the ExtremeXP project (EU Horizon program, GA 101093164).

% \begin{acks}
% This work was supported by the ExtremeXP project (EU Horizon program, GA 101093164).
% \end{acks}

% \footnotesize
% \small
% % \bibliographystyle{IEEEtran}
\bibliographystyle{ACM-Reference-Format}

\begin{thebibliography}{21}

%%% ====================================================================
%%% NOTE TO THE USER: you can override these defaults by providing
%%% customized versions of any of these macros before the \bibliography
%%% command.  Each of them MUST provide its own final punctuation,
%%% except for \shownote{}, \showDOI{}, and \showURL{}.  The latter two
%%% do not use final punctuation, in order to avoid confusing it with
%%% the Web address.
%%%
%%% To suppress output of a particular field, define its macro to expand
%%% to an empty string, or better, \unskip, like this:
%%%
%%% \newcommand{\showDOI}[1]{\unskip}   % LaTeX syntax
%%%
%%% \def \showDOI #1{\unskip}           % plain TeX syntax
%%%
%%% ====================================================================

\ifx \showCODEN    \undefined \def \showCODEN     #1{\unskip}     \fi
\ifx \showDOI      \undefined \def \showDOI       #1{#1}\fi
\ifx \showISBNx    \undefined \def \showISBNx     #1{\unskip}     \fi
\ifx \showISBNxiii \undefined \def \showISBNxiii  #1{\unskip}     \fi
\ifx \showISSN     \undefined \def \showISSN      #1{\unskip}     \fi
\ifx \showLCCN     \undefined \def \showLCCN      #1{\unskip}     \fi
\ifx \shownote     \undefined \def \shownote      #1{#1}          \fi
\ifx \showarticletitle \undefined \def \showarticletitle #1{#1}   \fi
\ifx \showURL      \undefined \def \showURL       {\relax}        \fi
% The following commands are used for tagged output and should be
% invisible to TeX
\providecommand\bibfield[2]{#2}
\providecommand\bibinfo[2]{#2}
\providecommand\natexlab[1]{#1}
\providecommand\showeprint[2][]{arXiv:#2}

\bibitem[\protect\citeauthoryear{Agarwal, Mozafari, Panda, Milner, Madden, and
  Stoica}{Agarwal et~al\mbox{.}}{2013}]%
        {AgarwalMPMMS13}
\bibfield{author}{\bibinfo{person}{Sameer Agarwal}, \bibinfo{person}{Barzan
  Mozafari}, \bibinfo{person}{Aurojit Panda}, \bibinfo{person}{Henry Milner},
  \bibinfo{person}{Samuel Madden}, {and} \bibinfo{person}{Ion Stoica}.}
  \bibinfo{year}{2013}\natexlab{}.
\newblock \showarticletitle{{B}linkdb: {Q}ueries with {B}ounded {E}rrors and
  {B}ounded {R}esponse {T}imes on {V}ery {L}arge {D}ata}. In
  \bibinfo{booktitle}{\emph{EuroSys}}.
\newblock


\bibitem[\protect\citeauthoryear{Alagiannis, Borovica, Branco, Idreos, and
  Ailamaki}{Alagiannis et~al\mbox{.}}{2012}]%
        {Alagiannis2012}
\bibfield{author}{\bibinfo{person}{Ioannis Alagiannis}, \bibinfo{person}{Renata
  Borovica}, \bibinfo{person}{Miguel Branco}, \bibinfo{person}{Stratos Idreos},
  {and} \bibinfo{person}{Anastasia Ailamaki}.} \bibinfo{year}{2012}\natexlab{}.
\newblock \showarticletitle{{N}odb: {E}fficient {Q}uery {E}xecution on {R}aw
  {D}ata {F}iles}. In \bibinfo{booktitle}{\emph{SIGMOD}}.
\newblock


\bibitem[\protect\citeauthoryear{Bikakis, Maroulis, Papastefanatos, and
  Vassiliadis}{Bikakis et~al\mbox{.}}{2021}]%
        {IS}
\bibfield{author}{\bibinfo{person}{Nikos Bikakis}, \bibinfo{person}{Stavros
  Maroulis}, \bibinfo{person}{George Papastefanatos}, {and}
  \bibinfo{person}{Panos Vassiliadis}.} \bibinfo{year}{2021}\natexlab{}.
\newblock \showarticletitle{{I}n-situ {V}isual {E}xploration over {B}ig {R}aw
  {D}ata}.
\newblock \bibinfo{journal}{\emph{Information Systems}} (\bibinfo{year}{2021}).
\newblock


\bibitem[\protect\citeauthoryear{Ghosh, Eldawy, and Jais}{Ghosh
  et~al\mbox{.}}{2019}]%
        {GhoshEJ19}
\bibfield{author}{\bibinfo{person}{Saheli Ghosh}, \bibinfo{person}{Ahmed
  Eldawy}, {and} \bibinfo{person}{Shipra Jais}.}
  \bibinfo{year}{2019}\natexlab{}.
\newblock \showarticletitle{{AID:} An Adaptive Image Data Index for Interactive
  Multilevel Visualization}. In \bibinfo{booktitle}{\emph{{ICDE}}}.
\newblock


\bibitem[\protect\citeauthoryear{Holanda and Manegold}{Holanda and
  Manegold}{2021}]%
        {HolandaM21}
\bibfield{author}{\bibinfo{person}{Pedro Holanda} {and} \bibinfo{person}{Stefan
  Manegold}.} \bibinfo{year}{2021}\natexlab{}.
\newblock \showarticletitle{Progressive Mergesort: Merging Batches of Appends
  into Progressive Indexes}. In \bibinfo{booktitle}{\emph{EDBT}}.
\newblock


\bibitem[\protect\citeauthoryear{Kim, Blais, Parameswaran, Indyk, Madden, and
  Rubinfeld}{Kim et~al\mbox{.}}{2015}]%
        {KimBPIMR15}
\bibfield{author}{\bibinfo{person}{Albert Kim}, \bibinfo{person}{Eric Blais},
  \bibinfo{person}{Aditya~G. Parameswaran}, \bibinfo{person}{Piotr Indyk},
  \bibinfo{person}{Samuel Madden}, {and} \bibinfo{person}{Ronitt Rubinfeld}.}
  \bibinfo{year}{2015}\natexlab{}.
\newblock \showarticletitle{Rapid Sampling for Visualizations with Ordering
  Guarantees}.
\newblock \bibinfo{journal}{\emph{PVLDB}} (\bibinfo{year}{2015}).
\newblock


\bibitem[\protect\citeauthoryear{Lampropoulos, Zardbani, Mamoulis, and
  Karras}{Lampropoulos et~al\mbox{.}}{2023}]%
        {0002ZMK23}
\bibfield{author}{\bibinfo{person}{Konstantinos Lampropoulos},
  \bibinfo{person}{Fatemeh Zardbani}, \bibinfo{person}{Nikos Mamoulis}, {and}
  \bibinfo{person}{Panagiotis Karras}.} \bibinfo{year}{2023}\natexlab{}.
\newblock \showarticletitle{Adaptive Indexing in High-Dimensional Metric
  Spaces}.
\newblock \bibinfo{journal}{\emph{PVLDB}} (\bibinfo{year}{2023}).
\newblock


\bibitem[\protect\citeauthoryear{Liang, Sintos, Shang, and Krishnan}{Liang
  et~al\mbox{.}}{2021}]%
        {Liang21}
\bibfield{author}{\bibinfo{person}{Xi Liang}, \bibinfo{person}{Stavros Sintos},
  \bibinfo{person}{Zechao Shang}, {and} \bibinfo{person}{Sanjay Krishnan}.}
  \bibinfo{year}{2021}\natexlab{}.
\newblock \showarticletitle{Combining Aggregation and Sampling (Nearly)
  Optimally for Approximate Query Processing}. In
  \bibinfo{booktitle}{\emph{SIGMOD}}.
\newblock


\bibitem[\protect\citeauthoryear{Liu, Jiang, and Heer}{Liu
  et~al\mbox{.}}{2013}]%
        {liu2013immens}
\bibfield{author}{\bibinfo{person}{Zhicheng Liu}, \bibinfo{person}{Biye Jiang},
  {and} \bibinfo{person}{Jeffrey Heer}.} \bibinfo{year}{2013}\natexlab{}.
\newblock \showarticletitle{imMens: Real-time visual querying of big data}. In
  \bibinfo{booktitle}{\emph{Computer graphics forum}}.
\newblock


\bibitem[\protect\citeauthoryear{Maroulis, Bikakis, Papastefanatos,
  Vassiliadis, and Vassiliou}{Maroulis et~al\mbox{.}}{2021}]%
        {MaroulisBPVV21}
\bibfield{author}{\bibinfo{person}{Stavros Maroulis}, \bibinfo{person}{Nikos
  Bikakis}, \bibinfo{person}{George Papastefanatos}, \bibinfo{person}{Panos
  Vassiliadis}, {and} \bibinfo{person}{Yannis Vassiliou}.}
  \bibinfo{year}{2021}\natexlab{}.
\newblock \showarticletitle{RawVis: {A} System for Efficient In-situ Visual
  Analytics}. In \bibinfo{booktitle}{\emph{SIGMOD}}.
\newblock


\bibitem[\protect\citeauthoryear{Maroulis, Bikakis, Papastefanatos,
  Vassiliadis, and Vassiliou}{Maroulis et~al\mbox{.}}{2022}]%
        {MaroulisS22}
\bibfield{author}{\bibinfo{person}{Stavros Maroulis}, \bibinfo{person}{Nikos
  Bikakis}, \bibinfo{person}{George Papastefanatos}, \bibinfo{person}{Panos
  Vassiliadis}, {and} \bibinfo{person}{Yannis Vassiliou}.}
  \bibinfo{year}{2022}\natexlab{}.
\newblock \showarticletitle{Resource-Aware Adaptive Indexing for In-situ Visual
  Exploration and Analytics}.
\newblock \bibinfo{journal}{\emph{VLDBJ}} (\bibinfo{year}{2022}).
\newblock


\bibitem[\protect\citeauthoryear{Maroulis, Stamatopoulos, Papastefanatos, and
  Terrovitis}{Maroulis et~al\mbox{.}}{2024}]%
        {MaroulisVLDB24}
\bibfield{author}{\bibinfo{person}{Stavros Maroulis}, \bibinfo{person}{Vassilis
  Stamatopoulos}, \bibinfo{person}{George Papastefanatos}, {and}
  \bibinfo{person}{Manolis Terrovitis}.} \bibinfo{year}{2024}\natexlab{}.
\newblock \showarticletitle{Visualization-Aware Time Series Min-Max Caching
  with Error Bound Guarantees}.
\newblock \bibinfo{journal}{\emph{PVLDB}} (\bibinfo{year}{2024}).
\newblock


\bibitem[\protect\citeauthoryear{Olma, Karpathiotakis, Alagiannis,
  Athanassoulis, and Ailamaki}{Olma et~al\mbox{.}}{2020}]%
        {Olma20}
\bibfield{author}{\bibinfo{person}{Matthaios Olma}, \bibinfo{person}{Manos
  Karpathiotakis}, \bibinfo{person}{Ioannis Alagiannis}, \bibinfo{person}{Manos
  Athanassoulis}, {and} \bibinfo{person}{Anastasia Ailamaki}.}
  \bibinfo{year}{2020}\natexlab{}.
\newblock \showarticletitle{Adaptive partitioning and indexing for in situ
  query processing}.
\newblock \bibinfo{journal}{\emph{The VLDB Journal}}  \bibinfo{volume}{29}
  (\bibinfo{date}{01} \bibinfo{year}{2020}).
\newblock


\bibitem[\protect\citeauthoryear{Park, Cafarella, and Mozafari}{Park
  et~al\mbox{.}}{2016}]%
        {ParkCM16}
\bibfield{author}{\bibinfo{person}{Yongjoo Park}, \bibinfo{person}{Michael~J.
  Cafarella}, {and} \bibinfo{person}{Barzan Mozafari}.}
  \bibinfo{year}{2016}\natexlab{}.
\newblock \showarticletitle{Visualization-aware sampling for very large
  databases}. In \bibinfo{booktitle}{\emph{ICDE}}.
\newblock


\bibitem[\protect\citeauthoryear{Peng, Zhang, Wang, and Pei}{Peng
  et~al\mbox{.}}{2018}]%
        {AQP++}
\bibfield{author}{\bibinfo{person}{Jinglin Peng}, \bibinfo{person}{Dongxiang
  Zhang}, \bibinfo{person}{Jiannan Wang}, {and} \bibinfo{person}{Jian Pei}.}
  \bibinfo{year}{2018}\natexlab{}.
\newblock \showarticletitle{AQP++: Connecting Approximate Query Processing With
  Aggregate Precomputation for Interactive Analytics}. In
  \bibinfo{booktitle}{\emph{SIGMOD}}.
\newblock


\bibitem[\protect\citeauthoryear{Rahman, Aliakbarpour, Kong, Blais, Karahalios,
  Parameswaran, and Rubinfeld}{Rahman et~al\mbox{.}}{2017}]%
        {RahmanAKBKPR17}
\bibfield{author}{\bibinfo{person}{Sajjadur Rahman}, \bibinfo{person}{Maryam
  Aliakbarpour}, \bibinfo{person}{Hidy Kong}, \bibinfo{person}{Eric Blais},
  \bibinfo{person}{Karrie Karahalios}, \bibinfo{person}{Aditya~G.
  Parameswaran}, {and} \bibinfo{person}{Ronitt Rubinfeld}.}
  \bibinfo{year}{2017}\natexlab{}.
\newblock \showarticletitle{{I}'ve {S}een "{E}nough": {I}ncrementally
  {I}mproving {V}isualizations to {S}upport {R}apid {D}ecision {M}aking}.
\newblock \bibinfo{journal}{\emph{PVLDB}} (\bibinfo{year}{2017}).
\newblock


\bibitem[\protect\citeauthoryear{Shang, Zgraggen, Buratti, Eichmann,
  Karimeddiny, Meyer, Runnels, and Kraska}{Shang et~al\mbox{.}}{2021}]%
        {Davos21}
\bibfield{author}{\bibinfo{person}{Zeyuan Shang}, \bibinfo{person}{Emanuel
  Zgraggen}, \bibinfo{person}{Benedetto Buratti}, \bibinfo{person}{Philipp
  Eichmann}, \bibinfo{person}{Navid Karimeddiny}, \bibinfo{person}{Charlie
  Meyer}, \bibinfo{person}{Wesley Runnels}, {and} \bibinfo{person}{Tim
  Kraska}.} \bibinfo{year}{2021}\natexlab{}.
\newblock \showarticletitle{Davos: a system for interactive data-driven
  decision making}.
\newblock \bibinfo{journal}{\emph{PVLDB}} (\bibinfo{year}{2021}).
\newblock


\bibitem[\protect\citeauthoryear{Ulmer, Angelini, Fekete, Kohlhammer, and
  May}{Ulmer et~al\mbox{.}}{2024}]%
        {hal-04361344}
\bibfield{author}{\bibinfo{person}{Alex Ulmer}, \bibinfo{person}{Marco
  Angelini}, \bibinfo{person}{Jean-Daniel Fekete}, \bibinfo{person}{J{\"o}rn
  Kohlhammer}, {and} \bibinfo{person}{Thorsten May}.}
  \bibinfo{year}{2024}\natexlab{}.
\newblock \showarticletitle{{A Survey on Progressive Visualization}}.
\newblock \bibinfo{journal}{\emph{{IEEE TVCG }}} (\bibinfo{year}{2024}).
\newblock


\bibitem[\protect\citeauthoryear{Wang, Wang, Chen, Zhao, Zhang, Wu, Fu, and
  Yu}{Wang et~al\mbox{.}}{2023}]%
        {OM323}
\bibfield{author}{\bibinfo{person}{Yunhai Wang}, \bibinfo{person}{Yuchun Wang},
  \bibinfo{person}{Xin Chen}, \bibinfo{person}{Yue Zhao}, \bibinfo{person}{Fan
  Zhang}, \bibinfo{person}{Eugene Wu}, \bibinfo{person}{Chi-Wing Fu}, {and}
  \bibinfo{person}{Xiaohui Yu}.} \bibinfo{year}{2023}\natexlab{}.
\newblock \showarticletitle{OM3: An Ordered Multi-level Min-Max Representation
  for Interactive Progressive Visualization of Time Series}.
\newblock \bibinfo{journal}{\emph{SIGMOD}} (\bibinfo{year}{2023}).
\newblock


\bibitem[\protect\citeauthoryear{Yu and Sarwat}{Yu and Sarwat}{2020}]%
        {0001S20}
\bibfield{author}{\bibinfo{person}{Jia Yu} {and} \bibinfo{person}{Mohamed
  Sarwat}.} \bibinfo{year}{2020}\natexlab{}.
\newblock \showarticletitle{Turbocharging Geospatial Visualization Dashboards
  via a Materialized Sampling Cube Approach}. In
  \bibinfo{booktitle}{\emph{{ICDE}}}.
\newblock


\bibitem[\protect\citeauthoryear{Zardbani, Mamoulis, Idreos, and
  Karras}{Zardbani et~al\mbox{.}}{2023}]%
        {ZardbaniMIK23}
\bibfield{author}{\bibinfo{person}{Fatemeh Zardbani}, \bibinfo{person}{Nikos
  Mamoulis}, \bibinfo{person}{Stratos Idreos}, {and}
  \bibinfo{person}{Panagiotis Karras}.} \bibinfo{year}{2023}\natexlab{}.
\newblock \showarticletitle{Adaptive Indexing of Objects with Spatial Extent}.
\newblock \bibinfo{journal}{\emph{PVLDB}} (\bibinfo{year}{2023}).
\newblock


\end{thebibliography}
%%% -*-BibTeX-*-
%%% Do NOT edit. File created by BibTeX with style
%%% ACM-Reference-Format-Journals [18-Jan-2012].

\end{document}